\documentclass{nature}

\usepackage{ulem}

\newcommand{\aap}{Astron.~Astrophys.}  
\newcommand{\apjl}{Astrophys.~J.~Letters}   
\newcommand{\apjs}{Astrophys.~J.~Supp.}   

\newcommand{\farcs}{\ensuremath{\overset{\prime\prime}{.}}}

\usepackage{graphicx}
\usepackage{natbib}
\usepackage{amsmath,amssymb,gensymb}
\usepackage{color}
\usepackage{bm}
\usepackage[utf8]{inputenc}

\title{Disk-mediated accretion burst in a high-mass young stellar object}

\author{A. Caratti o Garatti$^1$, B. Stecklum$^2$, R. Garcia Lopez$^1$, J. Eisl\"{o}ffel$^2$, T.P. Ray$^1$, A. Sanna$^3$, R. Cesaroni$^4$, C.M. Walmsley$^{1,4}$, R.D. Oudmaijer$^5$, W.J. de Wit$^6$,
L. Moscadelli$^4$, J. Greiner$^7$, A. Krabbe$^8$, C. Fischer$^8$, R. Klein$^9$, J.M. Iba\~{n}ez$^{10}$ }

\begin{document}

\maketitle
%
\begin{affiliations}
 \item Dublin Institute for Advanced Studies, Astronomy \& Astrophysics Section,
31 Fitzwilliam Place, Dublin 2, Ireland.
\item Th\"{u}ringer Landessternwarte Tautenburg, Sternwarte 5, D-07778 Tautenburg, Germany.
\item Max Planck Institut f\"ur Radioastronomie, Auf dem H\"{u}gel 69, D-53121 Bonn, Germany.
\item INAF – Osservatorio Astrofisico di Arcetri, Largo E. Fermi 5, 50125 Firenze, Italy.
\item School of Physics and Astronomy, University of Leeds, Leeds, LS2 9JT, UK.
\item ESO-European Organisation for Astronomical Research in the Southern Hemisphere, Alonso de Cordova 3107, Vitacura, Santiago de Chile, Chile.
\item Max-Planck Institut f\"{u}r Extraterrestrische Physik, D-85741 Garching, Germany.
\item Deutsches SOFIA Institut, Pfaffenwaldring 29, D-70569 Stuttgart, Germany.
\item NASA Ames Research Center, Moffett Field, CA 94035, United States.
\item Instituto de Astrof\'isica de Andaluc\'ia (CSIC), Glorieta de la Astronom\'ia 3, E-18008 Granada, Spain.

\end{affiliations}

\begin{abstract}
Solar-mass stars form via circumstellar disk accretion (disk-mediated accretion). Recent findings indicate that this process is likely episodic in the form of accretion bursts~\citep{audard}, 
possibly caused by disk fragmentation~\citep{krat,krumholz07,vorobyov}. 
Although it cannot be ruled out that high-mass young stellar objects (HMYSOs; $\bm{M}>$8\,M$_\odot$, $\bm{L_{bol}}>$5$\times$10$^3$\,L$_\odot$) arise from the coalescence of their low-mass brethren~\citep{tan}, 
latest results suggest that they more likely form via disks~\citep{kraus,caratti16,kuiper,klassen16}.
Accordingly, disk-mediated accretion bursts should occur~\citep{tanmckee,krumholz}. Here we report on the discovery of the first disk-mediated accretion burst from a $\sim$20\,M$_\odot$ HMYSO~\citep{stecklum}. 
Our near-infrared images show the brightening of the central source and its outflow cavities. Near-infrared spectroscopy reveals emission lines typical of accretion bursts in low-mass protostars, 
but orders of magnitude more luminous. 
Moreover, the energy released  and the inferred mass-accretion rate are also orders of magnitude larger. Our results identify disk accretion as the common mechanism of star formation across 
the entire stellar mass spectrum.

\end{abstract}

S255IR\,NIRS\,3 (aka S255IR-SMA1) is a well studied $\sim$20\,M$_\odot$ (L$_{bol} \sim$ 2.4$\times$10$^4$\,L$_\odot$) HMYSO~\citep{zin,wang}  in the S255IR massive star forming region~\citep{zin},
located at a distance of $\sim$1.8\,kpc~\citep{burns}.
%
It displays a disk-like rotating structure~\citep{zin},
very likely an accretion disk, viewed nearly edge-on~\citep{boley} (inclination angle $\sim$80$^\circ$).
A molecular outflow has been detected~\citep{zin} (blue-shifted lobe position angle - P.A. - $\sim$247$^\circ$) perpendicular to the 
disk.
Two bipolar lobes (cavities), cleared by the outflow, are illuminated by the
central source and show up as reflection nebulae towards the SW (blue-shifted lobe) and NE (red-shifted lobe, see Figure~\ref{fig:Kband}, left panel). 
At $\sim$2\farcs5 west of NIRS\,3, another HMYSO, NIRS\,1 (aka S255IR-SMA2; M$_*\sim$8\,M$_\odot$~\citep{wang}), is also seen in the near-infrared.

Following the detection of a 6.7\,GHz class II methanol maser flare in the S255IR star-forming region~\citep{fujisawa},
we performed near-infrared imaging with the Panoramic Near Infrared Camera (PANIC) at the Calar Alto Observatory in November 2015 (see Methods), to check whether the flare was 
triggered by an accretion burst from one of the massive protostars in the region~\citep{stecklum}. Indeed, IR radiation from heated dust emitting at $\sim$20--30\,$\mu$m is thought 
to be the pumping mechanism of this maser transition~\citep{sobolev}.

Our images in the $H$ (1.65\,$\mu$m) and $Ks$ (2.16\,$\mu$m) bands reveal an increase in the IR brightness (burst) of S255IR\,NIRS\,3, 
by $\Delta H \sim$3.5\,mag and $\Delta K \sim$2.5\,mag with respect to the latest archival images taken with the UKIRT Infrared Deep Sky Survey (UKIDSS) in December 2009
(see Figure~\ref{fig:Kband}, upper left and upper right panels). Moreover, a substantial increase in brightness is also observed in the bipolar outflow cavities, which scatter the light from the central accreting source.
These findings provide evidence of an accretion burst
onto the HMYSO.
The lower left panel of Figure~\ref{fig:Kband} shows the brightness ratio between the first PANIC $Ks$-band image and the UKIDSS $K$-band frame (see Methods). The relative brightness distribution displays a bipolar appearance. 
In principle this effect could be the result of enhanced scattering in the outflow lobes or extinction variability. The former would require, however, an increase in the number density of grains by an order of magnitude
which is impossible to obtain within the short time between the UKIDSS and PANIC images.
Extinction variability is also excluded by our multi-wavelength observations which include near-, mid-, and far-infrared spectroscopy and imaging (see Methods).
Therefore, the only explanation for this phenomenon is
that we are observing the 
light from the burst scattered by the dust in the outflow cavities (the so called light echo).
Indeed, subsequent PANIC imaging
confirms this hypothesis by verifying the motion of the light echo, between November 2015 and February 2016 (see Fig.~\ref{fig:Kband}, lower right panel), as it moves away from the source. 
This discovery allows us to approximately date the onset of the burst around mid-June 2015 (see Methods).
Remarkably, this is the first light echo ever observed from the outburst of a high-mass young star.

The SINFONI/VLT $K$-band spectrum of NIRS\,3 (see Figure~\ref{fig:spectra}, left panel, black spectrum), obtained on the 26th of February 2016, shows a very red and almost featureless continuum, much brighter 
than that observed with the same instrument in the quiescent phase in 2007~\citep{wang} (see Figure~\ref{fig:spectra}, left panel, red spectrum). Notably, no photospheric features in absorption are detected.
The lack of prominent features and the extremely reddened continuum are likely due to: i) the strong veiling, caused by the accretion; ii)
the high visual extinction ($A_V$=44$\pm$16\,mag, see Methods), 
resulting from the large inclination of the circumstellar disk to our line of sight; and iii) the presence of a thick envelope surrounding the HMYSO. 

In contrast, $K$-band integral-field spectroscopy of the red-shifted lobe
(the brightest outflow cavity, see Fig.\ref{fig:Kband})
performed with SINFONI/VLT (March 2016) and NIFS/Gemini (April 2016),
reveals a wealth of spectral features from the burst (see Figure~\ref{fig:spectra}, right panel, black spectrum).
Indeed, the visual extinction towards the lobes (A$_V\!\sim$18$\pm$5\,mag and 28$\pm$9\,mag, blue- and red-shifted lobe, respectively; see Methods) is smaller than that towards the HMYSO itself.
The walls of the outflow cavities are acting as a mirror, scattering the light from the outbursting young star
and allowing us to peer directly into the central accretion region.
The right panel of Figure~\ref{fig:spectra} compares our NIFS/Gemini spectrum (in black) of the red-shifted outflow lobe with
the SINFONI/VLT pre-outburst spectrum~\citep{wang} of the same region (red spectrum). 
The new spectrum shows an increase in
luminosity for both the continuum and lines (H$_2$, Br$\gamma$), as well as the appearance of new emission lines, namely CO band-heads, Na\,I, He\,I, which are typically 
observed in young eruptive low-mass stars (EXors, FUors, and MNors~\citep{lorenzetti09,kospal,caratti13,audard,reipurth,contreras}) 
and are the typical signature of accretion disks, accretion and ejection activity.

Young eruptive low-mass stars (M$_*\lesssim$2\,M$_\odot$)
 of these groups produce accretion bursts
lasting from a few weeks up to decades, and with accretion luminosities up to thousands of solar luminosities~\citep{audard}. 
During the burst the mass accretion rate ($\dot{M}_{acc}$) usually increases from one (EXors) to several (MNors, FUors) orders of magnitude with respect to quiescence~\citep{audard,contreras}.
CO band-heads and Na\,I lines originate from the outer layer of the disk (within 1\,au from the central source in low-mass YSOs).
The inner disk atmosphere is heated up to temperatures of a few thousands Kelvin (2000--4000\,K) by the accretion burst~\citep{audard} and these lines show up in emission.
In contrast, both He\,I and Br$\gamma$ lines are emitted closer to the central source ($\lesssim$0.1\,au) and may originate from accretion onto the star and/or from disk winds~\citep{kospal}. 

The total luminosities of the Br$\gamma$ (2.2\,L$_\odot$), He\,I (1.2\,L$_\odot$), Na\,I (0.7\,L$_\odot$) and CO$_{v=2-0}$ (22\,L$_\odot$) lines during the burst of NIRS\,3 are from 
three to four orders of magnitude larger than those observed in EXors and MNors.
Drawing a parallel between high-mass and low-mass YSOs, this evidence is
suggesting that the size of the disk emitting region as well as the energy released by the burst are much larger in the present case.
Indeed, the luminosity derived from the spectral energy distribution (SED) of NIRS\,3 (see Figure\,\ref{fig:sed}) grows from (2.9$\pm^{1}_{0.7}$)$\times$10$^4$\,L$_\odot$ (red data points) 
to (1.6$\pm^{0.4}_{0.3}$)$\times$10$^5$\,L$_\odot$ (blue data points) during the burst (PANIC, GROND, VLT/SINFONI, SOFIA/FORCAST and FIFI-LS data), corresponding to an increase in accretion luminosity ($\Delta L_{acc}$) of 
(1.3$\pm^{0.4}_{0.3}$)$\times$10$^5$\,L$_\odot$
and an energy release of (1.2$\pm$0.4)$\times$10$^{46}$\,erg from the beginning of the burst until mid-April 2016 ($\sim$9 months, according to our latest observations), when the source was still in burst.
This latter amounts to an accreted mass of about two Jupiter masses (namely $\sim$3.4$\times$10$^{-3}$\,M$_\odot$, see Methods).
The derived quantities are about four orders of magnitude larger than what is found in EXors and MNors making this the most luminous accretion burst ever detected in a YSO.
Moreover, assuming that the mass of the central source is $\sim$20\,M$_\odot$ and its radius is equal to 10\,R$_\odot$ (approximately the radius of a $\sim$20\,M$_\odot$ star on the zero age main sequence),
from $\Delta L_{acc}$ we infer that $\dot{M}_{acc}$ is boosted to (5$\pm$2)$\times$10$^{-3}$\,M$_\odot$\,yr$^{-1}$ (see Methods). 
The inferred value is likely a lower limit, as the radius of a massive protostar should be several times larger than that of a main sequence star~\citep{hosokawa,hosokawa10}.
Nevertheless, the inferred mass accretion rate of this HMYSO burst is at least three orders of magnitude higher than those of EXors and MNors.

The accretion burst discovered in S255IR\,NIRS\,3 
adds fundamental information to our understanding
of the high-mass star formation process.
Our observations finally confirm that HMYSOs form through accretion disks at high mass accretion rates. 
Moreover they also  provide an observational proof of episodic accretion, likely originating from disk fragmentation. 
Here timescale and energetics of the outburst are more consistent with disk fragmentation
rather than stellar merger~\citep{bally} (see Methods).

In this respect, high-mass star formation can be considered as a scaled up version of the process by which low-mass stars are born. The main differences are that massive stars
would form through larger accretion disks with much higher mass accretion rates ($\geq$10$^{-4}$\,M$_\odot$\,yr$^{-1}$) and on shorter time scales. 


High mass accretion rates and the presence of an accretion disk are fundamental ingredients to circumvent the intense radiation pressure 
of the massive star, which otherwise might reduce and even halt accretion.
They allow further accretion to proceed even after the hydrogen burning starts~\citep{klassen16}.
At variance with low mass protostars, the timescale for gravitational contraction (Kelvin–Helmholtz time) is shorter than the timescale for accretion in HMYSOs,
producing a strong radiation field~\citep{hosokawa10}. 
The circumstellar disk reduces the radiation pressure allowing most of the radiation to escape through the bipolar cavities~\citep{krumholz05}.
Indeed the light echo and the increase in brightness of the outflow cavities in NIRS\,3 confirm this picture. 

Finally, as with low mass protostars, the accretion process would not be continuous but episodic.
This would also explain the observation of several knots in jets from HMYSOs~\citep{caratti}, assuming major outflows events can be linked to major accretion events. Indeed,
the morphology of different gas tracers along the outflow axis of
NIRS\,3, which shows a discrete number of knots,
suggests that the source experienced multiple
bursts within the last few thousand years~\citep{wang,zin,burns}.
Our burst detection proves the erratic behavior of the accretion process in HMYSOs. 
%
%
Indeed, several radiation hydrodynamic simulations predict the onset of accretion variability in high-mass star formation~\citep{kuiper,krumholz,klassen16}.
Notably, episodic accretion might also play an important role in regulating the ionizing radiation, bloating the central source, and prolonging the accretion time during the Ultra-Compact {H\,{\sc ii} (UCH\,{\sc ii}) phase~\citep{hosokawa16}.

\begin{methods}
\label{sec:methods}


\subsection{Infrared imaging of the burst.}
\label{sec:irimaging}
Near-infrared imaging at various epochs was performed with PANIC~\citep{baumeister} at the Calar Alto 2.2-m telescope and the Gamma-Ray Burst Optical/Near-Infrared Detector (GROND)~\citep{greiner} at the La Silla 2.2-m telescope. 
Basic image processing was performed by the instrument teams using the corresponding data pipelines. The photometric calibration was done using the Two Micron All Sky Survey (2MASS) catalogue~\citep{2mass}. 
Although short detector integration times were applied for the $K\!s$ band, partial saturation of the bright target and field stars of similar brightness was unavoidable, 
in particular under good seeing conditions (our typical seeing was $\lesssim$1''). This was accounted for by a continuous extension of the linear fit between catalog and instrumental magnitudes with a parabola 
for the brightest objects. 
Flux densities for NIRS\,3 were generally derived using the APER procedure from the IDL Astronomy Library~\citep{landsman}, taking the local background into account. 
Mid-and far-infrared flux densities were obtained by performing target-of-opportunity observations with FORCAST~\citep{adams} and FIFI-LS~\citep{klein,klein2} aboard SOFIA (PI J. Eisl\"offel, proposal ID 04\_0047). FORCAST images 
were taken using narrow-band filters centered at 7.7, 11.1, 19.7, 31.5, and 37.1\,$\mu$m. The spectral windows for FIFI-LS were chosen to match the central wavelengths of the 
far-infrared AKARI filters, i.e., 60, 90, 140, and 160\,$\mu$m. The FIFI-LS spectral data cubes, calibrated by a FIFI-LS team member (C. F.), were collapsed and photometry was performed on the mean image.

The photometric values of the burst shown in Fig.~\ref{fig:sed}, along with their error, photometric aperture, instrument and observation date, are reported in Table~\ref{tab:photometry}.


\subsection{Light echo.}
\label{sec:light echo}
To cancel the influence of non-uniform extinction for the assessment of the change of the scattered light distribution due to the burst
and to compensate the decreasing surface brightness with growing distance from the source, a ratio image between PANIC $K\!s$ and UKIDSS $K$ frames was calculated. Before doing so, 
the PSFs of the $K$ frame was convolved with a proper kernel to match that of the $K\!s$ frame. The applied photometric scaling factor was derived from the corresponding zero points of the images. 
This turned out to be correct since the brightness ratio for field stars is in the order of unity. 
The resulting distribution has an asymmetric bipolar morphology. 
The asymmetry results from the inclination of the scattering cavities relative to the sky plane, leading to larger light distances for the blue-shifted lobe 
for a given propagation period and vice versa. 

The surfaces of scattered light of fixed travel time can be approximated as paraboloids with the star at the origin. Thereby, it can be shown that at the onset of an outburst (t=0), 
the size ratio between the back- and forward-scattering lobes is zero and increases to unity over time. 
Thus, an approximately equal extent of the scattering lobes of a YSO seen close to edge-on is only expected for steady-state illumination. 
Moreover, for the purpose of judging the lobe sizes, it must also be taken into account that forward
scattering dominates in the blue-shifted lobe while backward scattering is
prevalent in the red-shifted lobe. Because of the different scattering
efficiencies, the red-shifted lobe will be less bright in general, and thus
appear smaller for a given surface brightness sensitivity.

The same analysis on a later PANIC $K\!s$ image (February 2016) confirms the light echo by verifying both its propagation and dilution. For deriving the onset of the burst (June 2015)
we estimated the light travel time derived from the mean of the extent of both lobes. 

\subsection{Spectral energy distribution.}
\label{sec:sed}
As the energy released by the burst is thermalized by dust grains and radiated away in the infrared, pre-burst fluxes in this wavelength range are crucial for deriving the increase in luminosity. 
For this purpose, pre-outburst non-saturated IRAC images of S255IR (taken in sub-array mode, courtesy of G. Fazio, program ID 40440) were retrieved from the IPAC infrared science archive. Image mosaics 
were obtained from the dithered images for each channel using a custom IDL\footnote{IDL is a trademark of Exelis Visual Information Solutions, Inc.} procedure. Flux densities for NIRS\,3 were estimated as described above. 
Similarly, flux densities for the N60 and N160 AKARI bands were derived from the corresponding images after retrieval from the ISAS/JAXA archive (the wide AKARI channels centered at 90 and 140\,$\mu$m are saturated). 
These data were complemented with an archival ISO/SWS spectrum (courtesy D. Whittet), $H$ and $K\!s$ VLT/ISAAC photometry (private communication by S. Correia, ESO proposal ID 074.C-0772(B)) as well as flux densities 
from the literature~\citep{wang,zin,longmore,itoh,simpson} and surveys (AKARI, BGPS, MSX, UKIDSS).
The outburst SED was obtained using data from PANIC, GROND, SINFONI, FORCAST and FIFI-LS taken in February 2016. The pre- and burst luminosities were derived by integrating the dereddened SEDs and assuming a distance 
of 1.8$\pm$0.1\,kpc. 
To deredden the SED we adopt our visual extinction $A_V$=\,44$\pm$16\,mag
and $R_V$\,=\,3.1 extinction law~\citep{draine}. The resulting pre- and outburst luminosities are 
(2.9$\pm^{1}_{0.7}$)$\times$10$^4$\,L$_\odot$ and  (1.6$\pm^{0.4}_{0.3}$)$\times$10$^5$\,L$_\odot$, respectively. The uncertainties
were inferred from the small distance error and the uncertainty on the visual extinction. 
We also note that because of the close to edge-on view of its circumstellar disk, the estimated luminosity might represent a lower limit. The proper value may be up to two times higher~\citep{whitney}.



\subsection{Infrared Integral Field Unit Spectroscopy.}

Our $K$-band (1.95--2.5\,$\mu$m) integral field unit (IFU) spectroscopic data of S255IR\,NIRS\,3 consist of three datasets
taken with SINFONI~\citep{eisenhauer} on VLT (ESO, Chile) with R$\sim$4000 and NIFS~\citep{mcgregor} on the Gemini North telescope with R$\sim$5300. Adaptive-optics assisted mode was used for all runs. 
The first SINFONI dataset (26th of February 2016) was centred on NIRS\,3 (25 mas pixel scale and field of view - FoV - of 0''.8$\times$0''.8).
The second SINFONI dataset (9th of March 2016) was taken with the lowest spatial sampling (250\,mas pixel scale and FoV of 8''$\times$8'') and maps an area of $\sim$11''$\times$11'' around NIRS\,3, covering NIRS\,3, NIRS\,1 
and their outflow cavities. 
NIFS data (100\,mas pixel scale and FoV of 3''$\times$3'') were collected on the 8th of April 2016 and maps the red-shifted outflow cavity covering an area of $\sim$6''$\times$6''.

SINFONI data were reduced with the standard reduction pipeline in GASGANO~\citep{modigliani} that includes dark and bad pixel removal, flat-field and optical distortion correction, wavelength calibration with arc lamps, 
and image combination to obtain the final 3D data cube.
NIFS data reduction was accomplished in a similar fashion using the Gemini package in IRAF. 

All data were corrected for atmospheric transmission and flux calibrated by means of standard stars.

SINFONI pre-outburst IFU spectra, taken between February and March 2007, were retrieved from the ESO Data Archive and already published in a previous paper~\citep{wang}. They map an area (70''$\times$70'') larger
than our observations. 
To compare pre- and outburst data, spectra were extracted from our data cubes within an area of 1''.5$\times$1.''5 (centred on NIRS\,3 source; $\rm RA(J2000):6^h12^m54.0^s; DEC(J2000):+17\degree59^{'}23.1^{''}$) and
6''$\times$6'' (centred on $RA(J200):6^h12^m54.4^s; DEC(J2000):+17\degree59^{'}24.7^{''}$) for NIRS\,3 (Figure~\ref{fig:spectra}, left panel) 
and the red-shifted outflow cavity (Figure~\ref{fig:spectra}, right panel), respectively.

\subsection{Visual extinction variability vs. accretion burst.}
\label{sec:Av}
In principle, large variations of the extinction towards NIRS\,3 could be 
a possible cause of the infrared variability of NIRS\,3. However, this argument does not fit our observations for the following reasons.

{\it a)} The increase in luminosity is detected at NIR, MIR and FIR wavelengths.
This implies that the variation in luminosity cannot be due to a change in visual extinction, that would indeed affect the NIR part of the SED but 
would just marginally affect the MIR part of the spectrum and would not affect its FIR portion. 
{\it b)} The increase in luminosity at IR wavelengths temporally matches the flares of the methanol masers in the radio.
Moreover the maser positions match that of NIRS 3 (Sanna et al., in preparation).
{\it c)} In addition, the light echo observed at NIR wavelengths matches the timing of the CH$_3$OH maser flares.
{\it d)} The increase in the SED luminosity matches the appearance (CO, He\,I, Na\,I, lines) and increase in luminosity (Br$\gamma$, H$_2$) of the IR lines.    
{\it e)} Visual extinction affects the intensity of both lines and continuum as well as the continuum's color.
As the extinction affects to the same extent both lines and continuum, the equivalent width (EW) of the lines should not change.
On the other hand, EWs and fluxes of Br$\gamma$ and H$_2$ lines, already present in the pre-outburst spectrum in the outflow cavity, show a large variability
and are anti-correlated, as expected in accretion events~\citep{lorenzetti13}.
This cannot be explained with extinction variability.
Moreover, the slope of the K-band spectra on source and outflow cavities does not show a significant change
before and during the outburst, i.e., we do not detect any blueing of the spectra in 2016.
{\it f)} As reported in the next subsection, the visual extinction towards the outflow cavities does not change significantly.

Therefore we infer that the visual extinction did not significantly change between 2007 and 2016. 


\subsection{Visual extinction towards the outflow cavities and on-source.}
To estimate the visual extinction towards both blue and red-shifted lobes, we use pairs of lines from [FeII]
(2.016/2.254\,$\mu$m) and H$_2$ (2.034/2.437\,$\mu$m, 2.122/2.424\,$\mu$m, 2.223/2.413\,$\mu$m)
species that originate from the same upper level. 
We detect shocked emission lines ([FeII] and H$_2$) in two knots positioned in the blue- and red-shifted lobes, respectively.
Assuming that the emission arises from optically thin gas, the observed line ratios depend only on the differential extinction. 
The theoretical values are derived from the Einstein coefficients~\citep{deb} 
and frequencies of the transitions. We adopt the Rieke \& Lebofsky~\citep{rieke} extinction
law to correct for the differential extinction and compute A$_V$.
Values inferred are A$_V$=18$\pm$5\,mag (A$_V$([FeII])=16$\pm$10\,mag and A$_V$(H$_2$)=19$\pm$5\,mag ) for the blue-shifted lobe and A$_V$=28$\pm$9\,mag (A$_V$([FeII])=27$\pm$14\,mag and A$_V$(H$_2$)=29$\pm$12\,mag) 
for the red-shifted lobe.
Similar values, but with larger uncertainties, are inferred from the pre-outburst spectra (2007) of the blue-shifted (A$_V$(H$_2$)=18$\pm$7\,mag) and red-shifted (A$_V$(H$_2$)=27$\pm$15\,mag) outflow cavities.
These latter measurements suggest that the visual extinction towards the lobes did not significantly change.

We also infer the visual extinction towards NIRS\,3 from the H$_2$ lines detected in the outburst spectrum (Fig.~\ref{fig:spectra}, left panel), obtaining A$_V$(H$_2$)=44$\pm$16\,mag.
The inferred value is consistent with A$_V$=46\,mag from Simpson et al. 2009~\citep{simpson}.
Finally, from the pre-outburst $J-H$ and $H-K$ colors of the UKIDSS photometry, we obtain A$_V\sim$48--62\,mag by assuming that NIRS\,3 is a O6 spectral type positioned on the ZAMS.
This latter is consistent with our previous estimate. 
Therefore we adopt A$_V$(H$_2$)=44$\pm$16\,mag towards the source and use this value to deredden the SED.

\subsection{Line Luminosity.}
The line luminosities in the red-shifted lobe were inferred from the deredden line fluxes using A$_V$=28$\pm$9\,mag and assuming a distance to the object of 1.8$\pm$0.1\,kpc~\citep{burns}. 

\subsection{Energy of burst, accreted mass and mass accretion rate.}
The burst energy ($E=\Delta L_{acc} \times \Delta t$, where $\Delta t$ is the length of the burst) delivered so far (until mid-April 2016, date of the last available observation) 
by the burst is inferred from $\Delta L_{acc}$=(1.3$\pm^{0.4}_{0.3}$)$\times$10$^5$\,L$_\odot$, obtained from the pre- and outburst SED, 
and considering that the burst begun around mid-June 2015. The accreted mass is inferred assuming that the stellar radius is $R_*$=10\,R$_\odot$ and using $E=GM_*M_{acc}/R_*$, where G is the gravitational constant, $M_*$ is 
the mass of the star and $M_{acc}$ is the accreted mass. Finally, the mass accretion rate is obtained from $\dot{M}_{acc}\sim(2 \Delta L_{acc} R_*)/(G M_*)$.

\subsection{Disk accretion by fragmentation vs merging.}

A conceptual question involves whether our observations can rule out the
possibility that what we are seeing is not disk fragmentation but
stellar capture and merger via tidal disruption~\citep{bally}. 
This scenario proposes that massive
stars build up by capturing other stars in disks, then tidally
disrupting them. However, both timescales
and energetics of the outburst of S255 NIRS\,3 seem to be
inconsistent with such a scenario.
For example, assuming that the mass of the central object is $\sim$20\,M$_\odot$,
the merger with a brown dwarf of 0.1\,M$_\odot$ would produce an energy of $\sim$5$\times$10$^{47}$\,erg released
in $\sim$10$^4$\,yr. These values are much larger than what we inferred from the outburst of NIRS\,3.

\subsection{Data availability.}

The datasets generated and analysed during the current study are not publicly available due to a proprietary period restriction of 12 months.
After this period ESO/SINFONI and GROND data will become publicly available from the European Southern Observatory science archive (http://archive.eso.org/eso/eso\_archive\_main.html)
under programs ID 296.C-5037(A) and 096.A-9099(A); Gemini/NIFS data from  the Gemini Observatory archive (https://archive.gemini.edu/searchform) under program ID GN-2016A-DD-5;
SOFIA/FORCAST and FIFI-LS data from the SOFIA science archive (https://dcs.sofia.usra.edu/dataRetrieval/SearchScienceArchiveInfoBasic.jsp) under program ID 04\_0047.
Upon request the authors will provide all data supporting this study.

\end{methods}



\begin{addendum}
 \item A.C.G., R.G.L., and T.P.R. were supported by Science Foundation Ireland, grant 13/ERC/I2907.
 A.S. was supported by the Deutsche
 Forschungsgemeinschaft (DFG) Priority Program 1573.
 We thank the ESO Paranal and Gemini Observatory staff for their support.
 B.S. thanks Sylvio Klose for helpful discussions concerning the light echo.
 This research is partly based on observations collected at the VLT (ESO Paranal, Chile) with programme 296.C-5037(A) and
 at the Gemini Observatory (Program ID GN-2016A-DD-5). Gemini Observatory 
 is operated by the Association of Universities for Research in Astronomy, Inc., under a cooperative agreement with the NSF 
 on behalf of the Gemini partnership: the National Science Foundation (United States), the National Research Council (Canada), CONICYT (Chile), 
 Ministerio de Ciencia, Tecnolog\'{i}a e Innovaci\'{o}n Productiva (Argentina), and Minist\'{e}rio da Ci\^{e}ncia, Tecnologia e Inova\c{c}\~{a}o (Brazil).
 \item[Competing Interests] The authors declare that they have no
competing financial interests.
 \item[Correspondence] Correspondence and requests for materials
should be addressed to Alessio Caratti o Garatti~(email: alessio@cp.dias.ie).
 \item[Author contributions]
A.C.G., B.S. and R.G.L. wrote the intial manuscript and worked on the data reduction and analysis.
R.G.L. supported SINFONI observations. 
A.C.G., B.S. and J.E. are the PIs of the ESO, Calar Alto as well as Gemini, and SOFIA proposals, respectively. 
A.S., R.C., and L.M. worked on the maser and radio data.
T.P.R, A.S., R.C., L.M., C.M.W., R.D.O., W.J.dW. are coauthors of the proposals.
J.G. provided GROND data. A.K., C.F., and R.K. supported SOFIA observations.
J.M.I. supported PANIC observations.
All coauthors commented the manuscript.
\end{addendum}

\begin{figure}
\centering
\includegraphics[width=15cm]{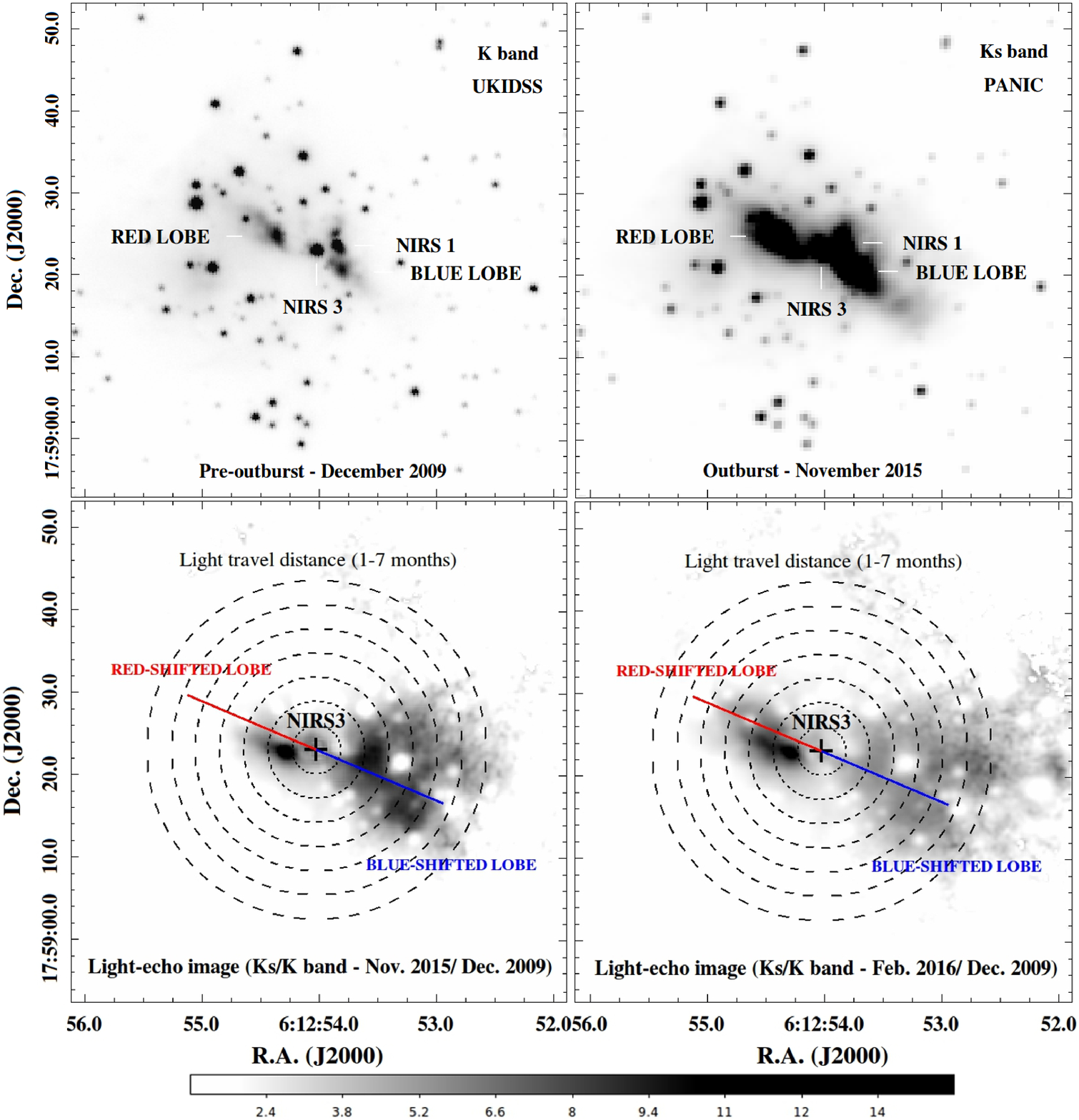}
\caption{
  {
  Pre-outburst, outburst and brightness-ratio images of S255IR\,NIRS\,3.
  {\bf [Upper left]} UKIDSS pre-outburst $K$-band image, December 2009: NIRS\,3 is centered on a bipolar nebula, towards north-east and south-west, namely the red- and blue-shifted outflow cavities of the protostar. 
  Another HMYSO (NIRS\,1) is situated $\sim$2.5'' west of NIRS\,3.
  {\bf [Upper right]} PANIC outburst $K\!s$-band image (November 2015), showing the brightening of NIRS\,3 and its outflow cavities. 
  {\bf [Lower left]} Ratio between PANIC $K\!s$ (Nov. 2015) and UKIDSS $K$ (Dec. 2009) images. 
  The gradual increment of brightness ratio towards the HMYSO represents the light echo -- a record of the burst history. 
  The echo asymmetry is primarily due to the outflow inclination with respect to the sky plane. For guidance concentric circles mark light travel distances in the plane of the sky separated by one month. 
  {\bf [Lower right] Ratio between PANIC $K\!s$ (Feb. 2016) and UKIDSS $K$ (Dec. 2009) images showing the
  motion of the light echo. The lower bar indicates the range of the relative brightness increase.}
  }
  \label{fig:Kband}
  }
\end{figure}

\begin{figure}
\centering
\includegraphics[width=6.5cm]{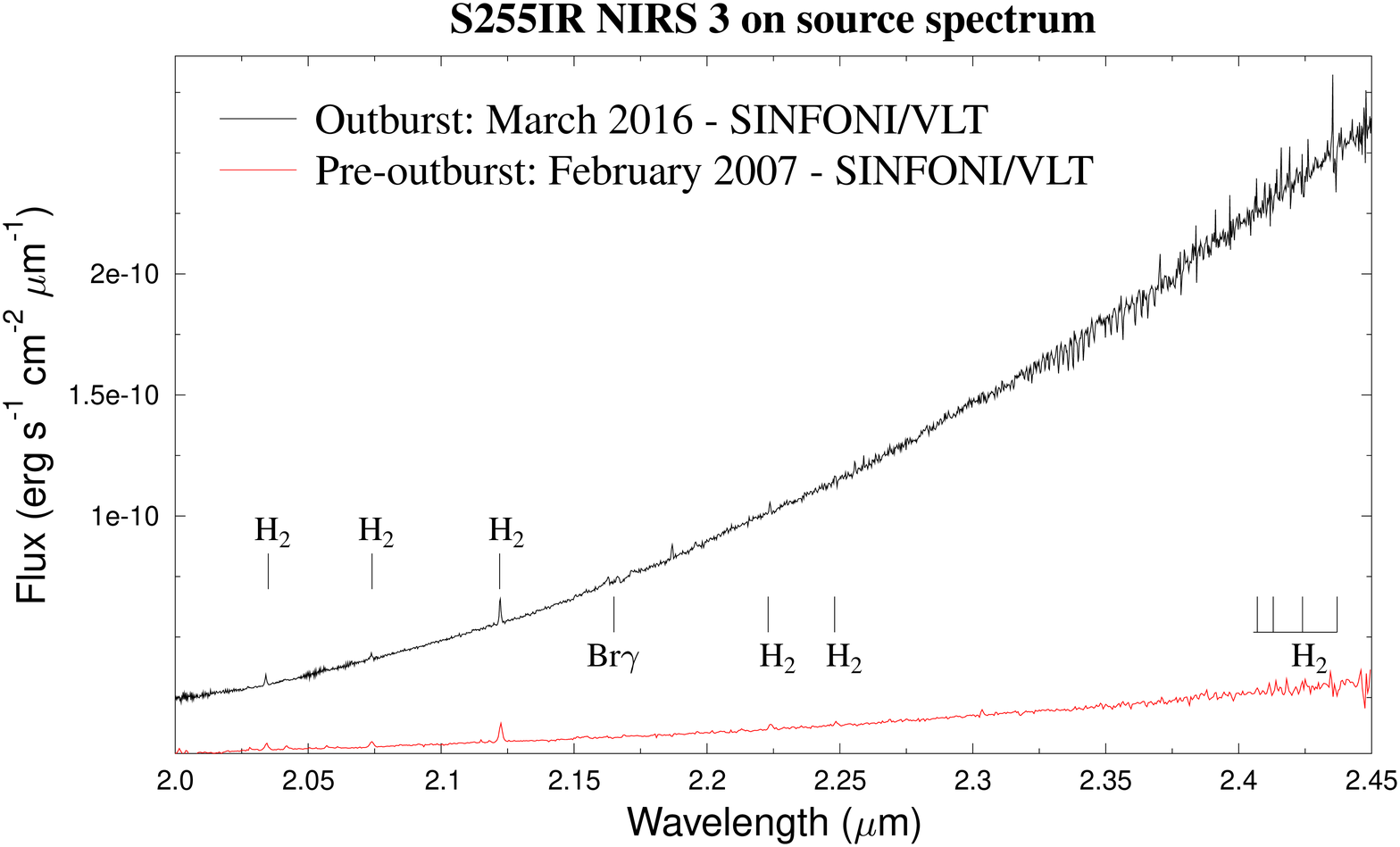}
\includegraphics[width=10.4cm]{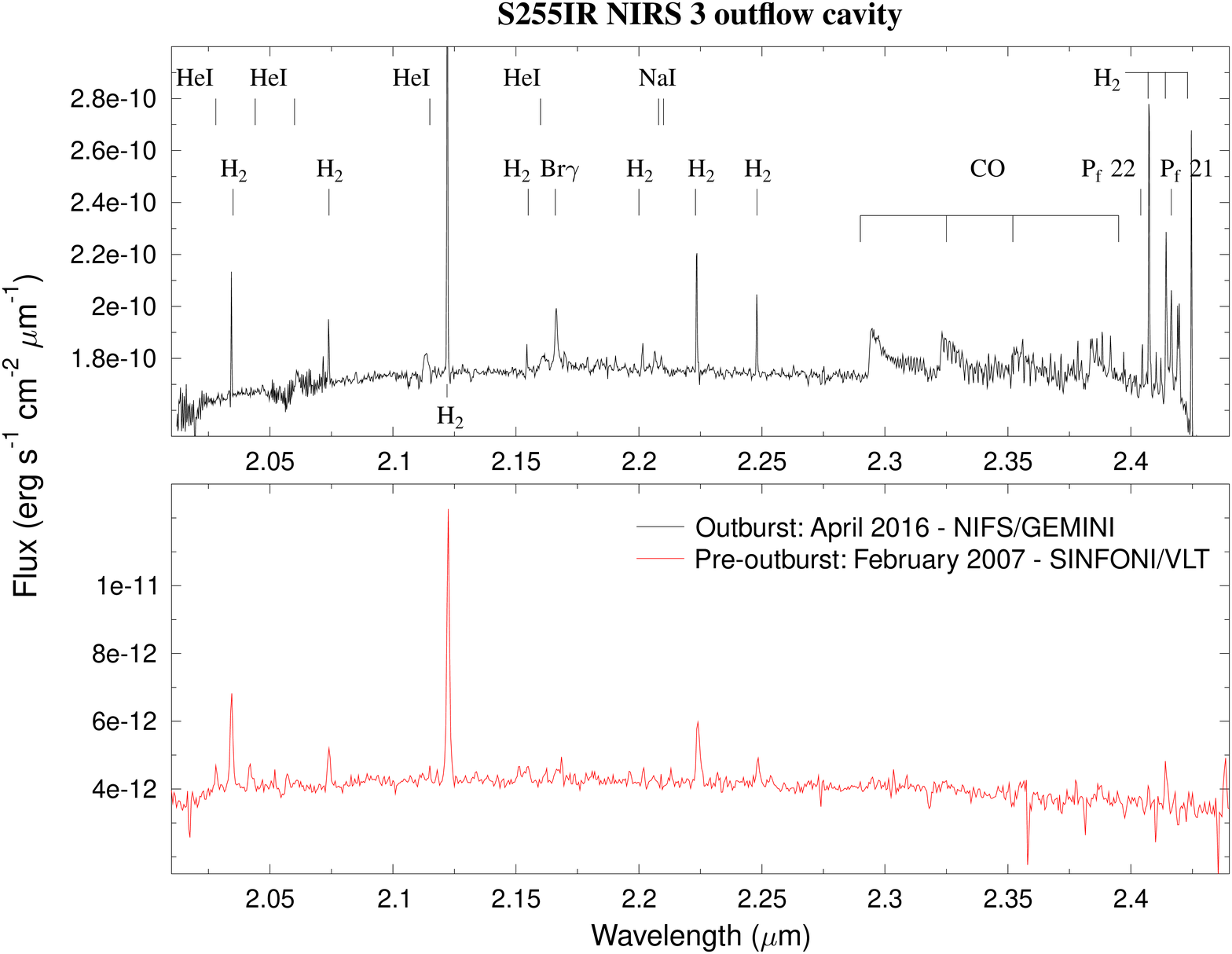}
\caption{
  { 
  Pre- and outburst K-band spectra of S255IR\,NIRS\,3 (left) and its red-shifted outflow cavity (right).
  {\bf [Left]} SINFONI/VLT pre-outburst (in red) and outburst (in black) K-band spectra of S255IR\,NIRS\,3.
  {\bf [Right]} SINFONI/VLT pre-outburst (in red) and outburst (in black) K-band spectra of the red-shifted outflow cavity of S255IR\,NIRS\,3. The spectrum in the outburst phase shows a large number 
  of emission lines typical of disk-mediated accretion outbursts.
  }
  \label{fig:spectra}
  }
\end{figure}

\begin{figure}
\centering
\includegraphics[width=11cm]{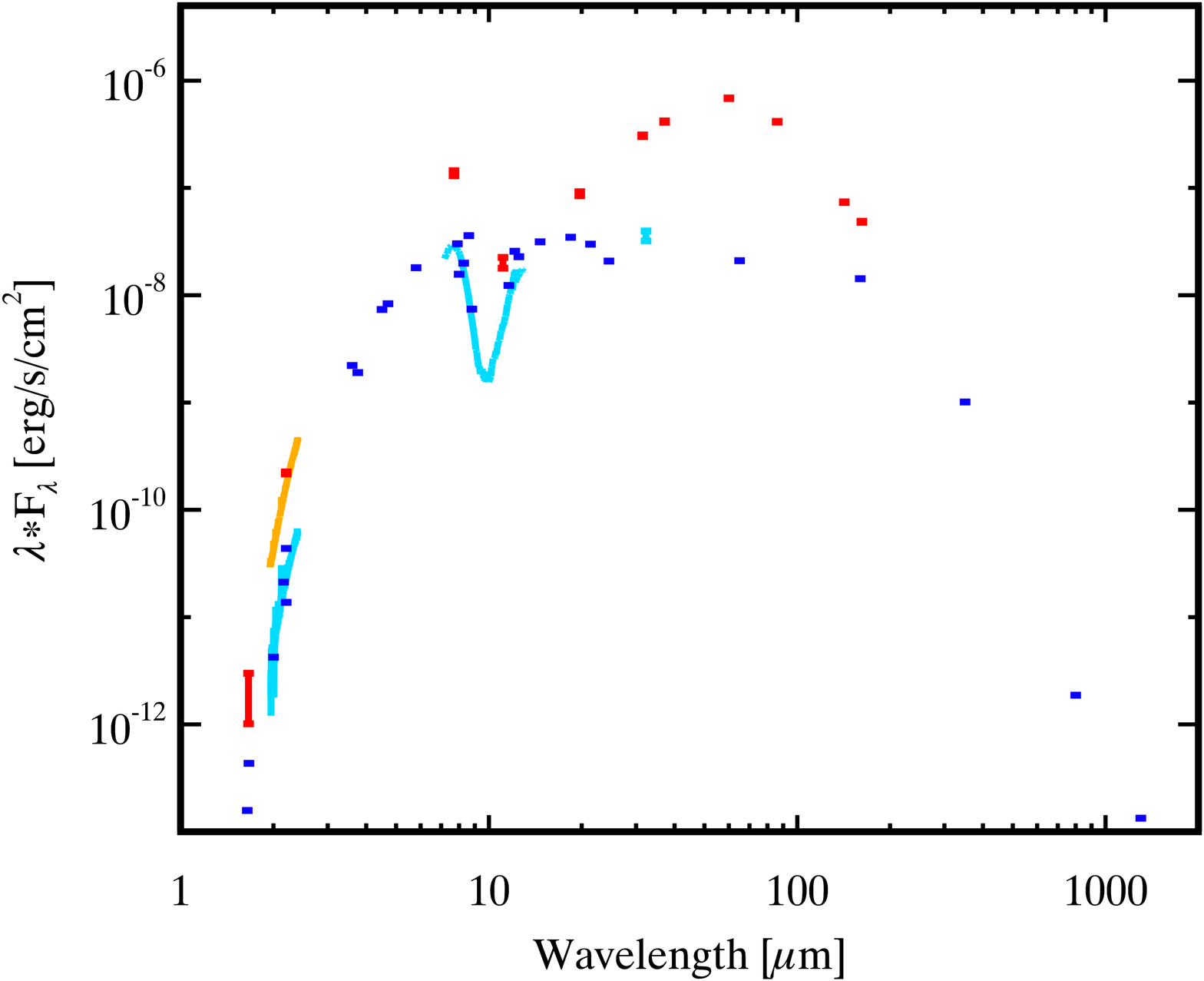}
\caption{
Pre- (cyan and blue) and outburst (orange and red) spectral energy distributions (SEDs) of S255IR\,NIRS\,3. Full colors indicate photometric measurements while light colors denote spectra. 

    \label{fig:sed}
  }
\end{figure}
\begin{table}
\caption{Flux densities of SED burst values}
\centering
\begin{tabular}{ c c c c c c }
  Wavelength & Flux density & Error & Aperture radius & Instrument & Date\\
  $\mu$m & [Jy] & [Jy] & [''] & & [D/M/Y] \\
  1.66  & 0.0011 &	0.0005 	& 0.6 	& PANIC & 15 01 2016\\
  2.2   & 0.1620 &	0.0032 	& 0.6 	& GROND	& 18 02 2016\\
  7.7   & 352.8	&	18.8 	& 5		& FORCAST & 04 02 2016\\
  11.1  & 74.8	&	8.6  	& 6  	& FORCAST & 04 02 2016\\
  19.7  & 580.9	&	24.1	& 8  	& FORCAST & 04 02 2016\\
  31.5  & 3223	&	56.8 	& 10 	& FORCAST & 04 02 2016\\
  37.1  & 5136	&	71.7 	& 11 	& FORCAST & 04 02 2016\\
  60	& 13720	&	97.2	& 10	& FIFI-LS &	01 03 2016\\
  86	& 11880	&	81.3	& 12	& FIFI-LS &	01 03 2016\\
 142	& 3490	&	24.2	& 20	& FIFI-LS &	01 03 2016\\
 162	& 2630	&	29.0	& 22	& FIFI-LS &	01 03 2016\\  
\end{tabular}
\label{tab:photometry}
\end{table}

\section{References}
\bibliographystyle{naturemag}
\bibliography{references}
\end{document}